\documentclass[aps,prc,twocolumn,showpacs,showkeys,nofootinbib,superscriptaddress]{revtex4-1}

\usepackage{longtable}
\usepackage{graphicx}
\usepackage{amsmath}
\usepackage{amsfonts}
\usepackage{amssymb}
\usepackage{natbib} 
\usepackage{setspace}
\usepackage{xspace}
\usepackage{cancel}
\usepackage{physics}
\usepackage{tensor,braket,color,bm,slashed}

\begin{document}

\title{
 \begin{flushright}
   \rightline{APCTP-17-23/7, LFTC-17-7/7} 
 \end{flushright}
Impact of medium modification of the nucleon weak and electromagnetic form factors 
on the neutrino mean free path in dense matter}

\author{Parada T. P. Hutauruk}
\email{parada.hutauruk@apctp.org}
\affiliation{Asia Pacific Center for Theoretical Physics, Pohang, Gyeongbuk 37673, Korea }

\author{Yongseok Oh}
\email{yohphy@knu.ac.kr}
\affiliation{Department of Physics, Kyungpook National University, Daegu 41566, Korea}
\affiliation{Asia Pacific Center for Theoretical Physics, Pohang, Gyeongbuk 37673, Korea }

\author{K. Tsushima}
\email{kazuo.tsushima@gmail.com}
\affiliation{Laborato\'rio de Fi\'sica Teo\'rica e Computacional, Universidade Cruzeiro 
do Sul, 01506-000, S\~ao Paulo, SP, Brazil }
\affiliation{Asia Pacific Center for Theoretical Physics, Pohang, Gyeongbuk 37673, Korea }

\date{\today}

\begin{abstract}
Impact of the in-medium modified nucleon weak and electromagnetic form factors on the
neutrino mean free path in dense matter is studied by considering both the weak and 
electromagnetic interactions of neutrinos with the constituents of the matter. 
A relativistic mean field model and the quark-meson coupling model are respectively 
adopted for the in-medium effective nucleon mass and nucleon form factors.
We find that the cross sections of neutrino scattering in cold nuclear medium 
decreases when the in-medium modification of the nucleon weak 
and electromagnetic form factors are taken into account. 
This reduction results in the enhancement of the neutrino mean free path, 
in particular at the baryon density of around a few times of the normal nuclear matter density. 
The enhancement of the neutrino mean free path is estimated to be about 10--40\% 
compared with the values obtained without the medium modification of the nucleon form factors, 
and the enhancement is expected to accelerate the cooling of neutron stars.   
\end{abstract}


\maketitle

\section{Introduction}
\label{intro}

It is widely accepted that the majority of neutrinos in the universe are produced in the core 
collapse supernova explosion. 
The final stage of the explosion creates a hot dense proto-neutron star, which 
emits bursts of neutrinos~\cite{BL86,BL87,RR16}. 
Then the produced neutrinos propagate through the neutron star and affect 
the evolution of neutron stars.
Inside the neutron star, neutrinos scatter with the constituents of matter, mostly 
neutrons and protons, and this process determines the propagation of neutrinos, namely,
the neutrino mean free path (NMFP).
Thus the NMFP is an important input in simulations of neutron star evolution 
as well as those of compact stars.
In previous calculations on the NMFP in neutrino scattering, it was found 
that the NMFP value for various neutrino scattering is larger than that of the neutrino absorption,
and the propagation of neutrino in neutron matter is longer than that in vacuum~\cite{SLVZ03}.

For estimating the neutrino scattering cross sections in a more realistic manner, 
various attempts were made by considering the effects of the phase space, weak magnetism, 
recoil correction, form factors, and strange quark corrections~\cite{GKLK93,HL99,Horowitz01,HP03}. 
In Ref.~\cite{SWHM06} the electromagnetic form factors of the neutrino were also considered.
These works, however, were based on the nucleon electromagnetic form factors in free space.

The electromagnetic form factors of nucleons reflect their internal structure.
For example, the anomalous magnetic moments of the nucleons and the momentum-dependence 
of the electromagnetic and axial form factors of nucleons are the consequences of 
their finite size, or their quark-gluon substructure.
Since not only the nucleons themselves but also the substructure of nucleons are expected to  
be modified in the surrounding environment, it is natural to expect that the electromagnetic
and weak properties of nucleons are also modified in nuclear medium. 
Thus, one of our main motivations of this study is to explore the impact of the possible 
in-medium modification of the nucleon electromagnetic and axial-vector form factors 
on the NMFP in nuclear medium. 
For this purpose, we first discuss the nucleon form factors in free space, 
and then elaborate the form factors in nuclear medium based on a relativistic phenomenological quark model,  
which is based on the quark degrees of freedom dictated by Quantum Chromodynamics (QCD).
Such an attempt was also recently made by the first principle lattice QCD calculations
for a few nucleon systems in Ref.~\cite{CDDG18}.

Although the detailed analysis is yet model-dependent, 
recent experimental observations in electron-nucleus scatterings suggest the in-medium 
modification of the nucleon electromagnetic (EM) 
form factors~\cite{JLab-A-00,DBBB00,Strauch03,CMPR09,E03-104-10,E03-104-11}.  
There are several issues related with the interpretation of the experimental observations in connection 
with the in-medium effects, nucleon correlations, and so on.
More detailed discussions can be found, for example, in Refs.~\cite{SBKMV04,Strauch12,BST11,HMPW16}.
Another example which may be interpreted as the in-medium modification of the nucleon properties 
would be the neutron life-time inside a nucleus. 
Namely, a bound neutron in a nucleus is found to live longer than in vacuum, where its life time in vacuum is 
$\simeq 880$ s~\cite{PDG16} against the weak beta-decay to the proton. 
This phenomenon would be interpreted as a change of axial-vector coupling constant, i.e.,
the effective axial-vector coupling constant in nuclear medium $g_A^{\rm eff}$ is smaller than 
that the in-vacuum value, $g_A^{} \simeq 1.27$. 
A recent review on the axial-vector coupling strength in $\beta$ and double $\beta$ decays can be found, for example,
in Ref.~\cite{Suhonen17}.
There are also alternative approaches to explain this observation while keeping 
the free space value of $g_A^{}$~\cite{PBCG17,EM16}.
Therefore, it is not easy to understand the correct origin of such observations, 
but what is clear is that many nontrivial contributions arise in nuclear medium, 
and the physics in nuclear medium is very different from the physics in free space. 
In the present article, following the point of view that the properties of the quark and gluon 
substructure of nucleons change in nuclear medium and
can be estimated by effective theories of QCD, 
we investigate the effects of in-medium modified weak and EM form factors 
of the nucleon on the NMFP in dense nuclear medium.

In previous studies of neutrino-nucleus scattering in Refs.~\cite{TKS03,CCKSKTM13},  
the authors, including one of us, estimated the (anti)-neutrino-nucleus scattering 
cross sections via charged~\cite{TKS03} and neutral~\cite{CCKSKTM13} 
currents for a bound nucleon by including the effect of the in-medium modified weak-magnetism 
and axial-vector form factors, 
$F_2^{W} (q^2)$ and $G_A (q^2)$, respectively, 
where $q$ is the transferred four-momentum.
(See e.g., Ref.~\cite{BHMM15} for a recent review on theoretical studies of 
neutrino-nucleus interactions.) 
They found that the in-medium modified nucleon form factors 
$F_2^{W} (q^2)$ and $G_A (q^2)$ as a total effect could reduce the total cross 
sections of neutrino-nucleus scattering by $\sim 8$\%~\cite{TKS03} and 12--18\%~\cite{CCKSKTM13}, respectively.
Motivated by these results, we address the role of the in-medium modified nucleon EM form factors 
in the present work by focusing on the impact of the medium modifications of both 
the weak and EM form factors of the nucleon in dense matter on the neutrino scattering with nucleons.

For this purpose, we adopt a relativistic mean field model inspired by the effective field theory 
models to describe nuclear matter. 
For the in-medium nucleon weak and EM form factors, we use 
the quark-meson coupling (QMC) model~\cite{Guichon88}. 
Based on these theoretical approaches, we calculate the differential and total cross sections 
of neutrino scatterings to estimate the NMFP.
We consider the scattering of neutrinos with the proton ($p$), the neutron ($n$), 
the electron ($e^-$), and the muon ($\mu^-$), 
which are the major constituents of the matter.
Our results may be used as a guidance for supernova simulations~\cite{MJBHMM15,MMHJ17}
to examine the effects of the in-medium nucleon form factors.

This paper is organized as follows. 
In Sec.~\ref{scattering}, we review the formalism for calculating cross sections 
of neutrino scatterings with the constituents of matter, i.e., nucleons, electrons, and muons.
In Sec.~\ref{mattermodels}, we briefly discuss the models of nuclear matter 
adopted in the present work. 
The in-medium modification of the weak and EM form factors of the nucleon 
are presented in Sec.~\ref{medium}. 
In Sec.~\ref{numerical}, our numerical results are presented and their implications are discussed. 
Section~\ref{summary} is devoted for a summary.

\section{Neutrino Scatterings with Matter Constituents}
\label{scattering}

In this section we calculate the differential cross sections of neutrino-matter scattering.  
Before discussing the in-medium modification of the nucleon weak and EM form 
factors, we briefly discuss the free space neutrino scatterings with constituents of the matter. 

The effective interaction of the neutrino with the constituents of matter is
given by the current-current interaction form as
\begin{eqnarray}
\label{eq:model1}
\mathcal{L}_{\rm int}^{j} &=& \frac{G_F}{\sqrt{2}} \left[ \bar{\nu}(k^{\prime}) 
\Gamma_{\rm W}^{\mu} \nu (k) \right] 
\left[ \bar{\psi}_j^{} (p^{\prime}) J_{\mu}^{{\rm W} (j)} \psi_j^{} (p) \right] 
\nonumber \\ && \mbox{}
+ \frac{4\, \alpha_{\rm em}^{}}{q^2} \left[ \bar{\nu}(k^{\prime}) 
\Gamma^{\mu}_{\rm EM} \nu (k) \right] 
\left[ \bar{\psi}_j^{} (p^{\prime}) J_{\mu}^{{\rm EM} (j)} \psi_j^{} (p) \right], 
\nonumber \\
\end{eqnarray}
where $G_F$ and $\alpha_{\rm em}^{}$ are the Fermi (weak)
coupling constant and the EM fine structure constant, respectively, whose values are
$G_F \simeq 1.166  \times 10^{-5}~\mbox{GeV}^{-2}$ and
$\alpha^{-1}_{\rm em} \simeq 137$~\cite{PDG16}.
The initial and final neutrino spinors are represented by $\nu (k)$ and $\bar{\nu} (k^{\prime})$, respectively, 
and $\psi_j^{} (p)$ and $\bar{\psi}_j^{} (p^{\prime})$ 
refer to the initial and final spinors of the target fermion $j$, respectively, where $j = (n,~p,~e^{-},~\mu^{-})$. 
The four-momenta of the initial and final neutrinos are denoted by $k$ and $k'$, respectively, and 
$p$ and $p'$ stand for the initial and final four-momenta of the target $j$.
The transferred four-momentum is thus $q=k- k^{\prime} = p^{\prime}-p$.
The first term of the Lagrangian in Eq.~(\ref{eq:model1}) is the current-current 
interaction between the neutrino and the nucleon. 
The weak interaction vertex of the neutrino is given by
\begin{align}
\label{eq:model2}
\Gamma_{\rm W}^{\mu} &= \gamma^{\mu} \left( 1 -\gamma^{5} \right).
\end{align}

The second term in the effective Lagrangian of Eq.~(\ref{eq:model1}) contains 
the current-current interaction of the EM interaction. 
Since the EM form factors of Majorana neutrinos can be obtained from those calculated for 
Dirac particles~\cite{Nieves82}, we consider the Dirac neutrino in the present work.
The EM vertex of Dirac neutrinos are described by four form 
factors as~\cite{Nieves82,VE89,GS08a,BGS12}%
\begin{eqnarray}
\label{eq:model3-1}
\Gamma^{\mu}_{\rm EM} (q^2) &=& f_{1} (q^2) \gamma^\mu - \frac{i}{2 m_e} f_{2 } (q^2) 
\sigma^{\mu \nu} q_{\nu}
\nonumber \\ && \mbox{}
+ g_{1}^{} (q^2) \left( g^{\mu \nu} - \frac{q^\mu q^\nu}{q^2} \right) \gamma_\nu^{} \gamma^5
\nonumber \\ && \mbox{}
- \frac{i}{2m_e} g_{2}^{} (q^2) \sigma^{\mu \nu} q_\nu^{} \gamma^5,
\end{eqnarray}
where the four form factors, $f_{1} (q^2)$, $g_{1}^{}(q^2)$, $f_{2} (q^2)$, and $g_{2}^{} (q^2)$, 
are the Dirac, anapole, magnetic, and electric form factors, respectively. 
Applying the current conservation condition, we can rewrite it as
\begin{eqnarray}
\label{eq:model3}
\Gamma_{\rm EM}^{\mu} (q^2) &=& 
f_{m}(q^2) \gamma^\mu + g_{1}^{} (q^2) \gamma^\mu \gamma^5, 
\nonumber \\ && \mbox{}
- \left[ f_{2}(q^2) + i g_{2}^{}(q^2) \gamma^5 \right] \frac{P^\mu}{2m_{e}^{}}, 
\end{eqnarray}
where $f_{m}(q^2) \equiv f_{1} (q^2) + ( m_\nu^{} / m_e^{} ) f_{2}(q^2)$ and $P_\mu = k_\mu + k_\mu^\prime$ 
with $m_\nu^{}$ ($m_{e}^{}$) being the neutrino (electron) mass.

In the static limit ($q^2=0$), the Dirac form factor $f_{1}(q^2)$ and the anapole 
form factor $g_{1}^{}(q^2)$ are related respectively to the vector charge radius $\braket{R_V^2}$ 
and the axial-vector charge radius 
$\braket{R_A^2}$ as~\cite{VE89,FS03,GS08a,BGS12} 
\begin{align}
\label{eq:model4}
\braket{R_V^2} &=  6 \left. \frac{d f_{1} (q^2)}{dq^2} \right \vert_{q^2 =0},  
\nonumber \\
\braket{R_A^2} &=  6 \left. \frac{d g_{1}^{} (q^2)}{dq^2} \right \vert_{q^2 =0},
\end{align}
which in the Breit frame, where $q_0^{} = 0$, leads to
\begin{align}
f_{1} (q^2) &\simeq \frac{1}{6} \braket{R_V^2} q^2
= - \frac{1}{6} \braket{R_V^2} \bm{q}^{\, 2}, 
\nonumber \\
g_{1}^{} (q^2) &\simeq \frac{1}{6} \braket{R_A^2} q^2 
= - \frac{1}{6} \braket{R_A^2} \bm{q}^{\, 2},
\end{align}
where we have used $f_{1}(0)=g_1^{}(0)=0$.
It is common to define the sum $\braket{R^2} \equiv  \braket{R_V^2} +  \braket{R_A^2}$, 
which can have a negative value for Dirac neutrinos~\cite{GKLLSZ15,Nardi03}. 
For Majorana neutrinos, only the anapole form factor $g_1^{}(q^2)$ remains
and thus the relevant quantity is solely $\braket{R_A^2}$~\cite{GS08a,Nieves82} 
in the EM interaction vertex with $\braket{R_{V}^2} = 0$~\cite{ GKLLSZ15,Nardi03}.
The values of $\abs{\braket{R^2}}$ for the Dirac neutrino are at the order of 
$(10^{-3}~\mbox{fm})^2$~\cite{CHARM-II-95,GKLLSZ15,Nardi03}.

At $q^2 = 0$, the form factors $f_{2}(q^2)$ and $g_{2}^{}(q^2)$ respectively define 
the neutrino magnetic moment and the charge parity (CP) violating electric dipole moment as
\begin{align}
\label{eq:model5}
\mu^{m} _\nu &= f_{2}(0) \mu_B^{} \qquad \mbox{and} \qquad
\mu_\nu^{e} = g_{2}^{} (0) \mu_B^{}, 
\end{align}
where the effective neutrino magnetic moment $\mu_\nu$ is defined as   
$\mu_\nu^2 \equiv \left( \mu_\nu^{m} \right)^2 + \left( \mu_\nu^{e}\right)^2$~\cite{KZAM92},  
and $\mu_B^{} = e/2m_e^{}$ is the Bohr magneton. 
The estimated value of the neutrino magnetic moment is around 
$10^{-10}\mu_B$~\cite{HSM06,SHM05b,HWSM04,KTAC90,Raffelt99,Borexino-17}.

For a free-space nucleon, the weak and EM vertices in the nucleon and lepton current operators  
read~\cite{,HSM06,KPH94,NPR04}
\begin{eqnarray}
\label{eq:model6}
J_{\mu}^{{\rm W}} &=& F_{1}^{{W}}(q^2) \gamma_\mu - G_{A}(q^2) \gamma_\mu \gamma^5, 
\nonumber \\ && \mbox{}
+ i F_{2}^{{W}}(q^2) \frac{\sigma_{\mu \nu} q^{\nu}}{2M_{N}} + 
\frac{G_p(q^2)}{2M_N}q_\mu \gamma^5,
\end{eqnarray}
and
\begin{eqnarray}
\label{eq:model7}
J_{\mu}^{{\rm EM}} = F_{1}^{{\rm EM}}(q^2) \gamma_\mu + i F_{2}^{{\rm EM}}(q^2) 
\frac{\sigma_{\mu\nu} q^\nu}{2M_N}
\end{eqnarray}
for each target particle $j$.
Since the induced pseudoscalar form factor $G_p (q^2)$ in Eq.~(\ref{eq:model6}) gives a very small 
contribution to the cross section, which is proportional to $\mbox{(lepton mass)}^2/M_N^2$~\cite{GS84},
it will be neglected in the present calculation.
The values of the weak form factors, $F_{1}^{W}(0)$, $G_A(0)$, and $F_{2}^{W}(0)$, 
in free space are listed in Table~\ref{tab:model1} with those of the EM form factors, 
$F_{1}^{\rm EM}(0)$ and $F_{2}^{\rm EM}(0)$.

\begin{table*}[t]
\caption{
Weak and electromagnetic form factor values at $q^2 = 0$ in free space. 
Here we use $\sin^2 \theta_w = 0.231$, $g_A^{} = 1.260$, 
$F^{\rm EM}_{2p}(0) \equiv \kappa_p^{} = 1.793$, and $F^{\rm EM}_{2n}(0) \equiv \kappa_n^{} = -1.913$ 
in the units of the nuclear magneton $\mu_N^{} = e/2M_p$ with $M_p$ being the proton mass. 
}
\label{tab:model1}
\addtolength{\tabcolsep}{1.2pt}
\begin{tabular}{cccccc} 
\hline \hline
\textrm{Target} & $F_1^{\rm W}$ & $G_A$ & $F_2^{\rm W}$ & $F_1^{\rm EM}$ & $F_2^{\rm EM}$ \\[0.2em] 
\hline
 $n$ & $-0.5$ &  $-\frac{g_A^{}}{2} $ & $-\frac{1}{2} \left( \kappa_p^{} - \kappa_n^{} \right) - 
 2  \sin^2 \theta_{w} \kappa_n^{}$ & 0 & $ \kappa_n^{}$ \\
 $p$ & $0.5-2 \sin^2 \theta_{w}$  &  $\frac{g_A^{}}{2} $ & 
 $\frac{1}{2} \left( \kappa_p^{} - \kappa_n^{} \right) - 2 \sin^2 \theta_{w} \kappa_n^{}$ & 1 
 & $\kappa_p^{}$ \\
 $e$ & $0.5+2 \sin^2 \theta_{w} $ &  $\frac{1}{2} $ & 0 & 1 & 0 \\
 $\mu$ & $-0.5+2 \sin^2 \theta_{w} $ & $- \frac{1}{2} $ & 0 & 1 & 0 
\\ \hline \hline
\end{tabular}
\end{table*}

With the effective Lagrangian of Eq.~(\ref{eq:model1}), the differential cross section per volume 
of the neutrino scattering with a target particle can be calculated as
\begin{widetext}
\begin{eqnarray}
\label{eq:model15}
\left( \frac{1}{V} \frac{d^3 \sigma}{d^2 \Omega^{'} dE_\nu^{'}} \right) 
&=& - \frac{1}{16 \pi^2} \frac{E_\nu^{'}}{E_\nu} 	\left[ \left( 
\frac{G_F}{\sqrt{2}}\right)^2 \left( L_{\nu}^{\alpha\beta} \Pi^{\rm Im}_{\alpha\beta} \right)^{(\rm W)}  
+ \left( \frac{4\pi \alpha_{\rm em}}{q^2} \right)^2 \left( L_{\nu}^{\alpha\beta} 
\Pi_{\alpha\beta}^{\rm Im} \right)^{(\rm EM)} 
+ \frac{8 \pi \,G_F \alpha_{\rm em}}{q^2 \sqrt{2}} \left( L_{\nu}^{\alpha\beta} 
\Pi_{\alpha\beta}^{\rm Im} \right)^{(\rm INT)} 	\right], 
\nonumber \\
\end{eqnarray}
\end{widetext}
where $E'_\nu$ ($E_{\nu}$) is the final (initial) energy of the neutrino. 
For the details on the analytic formulas of the  polarization tensors for the weak and 
EM interactions and all the corresponding quantities in Eq.~(\ref{eq:model15}),  
we refer to Refs.~\cite{SWHM06,SHM05b}. 
Then the final results of the contracted lepton and hadron tensors in the corresponding interactions  
in Eq.~(\ref{eq:model15}) are obtained as
\begin{eqnarray}
\label{eq:model16}
\left[ L_{\nu}^{\alpha \beta} \Pi_{\alpha \beta}^{\rm Im} \right]^{\rm (W)}  
&=& -8 q^2 \sum_{j} \left[ A_{W}^{j} \left( \Pi_{L}^{j} + \Pi_{T}^{j} \right) 
 + B_{ 1W}^{j} \Pi_{T}^{j} 
\right. \nonumber \\ &&  \left. \mbox{}
 + B_{ 2 W}^{j} \Pi_{A}^{j} + C_{ W}^{j} \Pi_{VA}^{j} \right], 
\nonumber \\
\label{eq:model17}
\left[ L_{\nu}^{\alpha \beta} \Pi_{\alpha \beta}^{\rm Im} \right]^{\rm (EM)}  
&=& q^2 \sum_{j} \left[ A_{\rm EM}^j \left( \Pi_{L}^{j} + \Pi_{T}^{j} 
\right) 
\right. \nonumber \\ && \left. \mbox{}
+ B_{\rm 1 EM}^{j} \Pi_{T}^{j} + B_{\rm 2 EM}^{j} \Pi_{A}^{j} \right],
\nonumber \\
\label{eq:model18}
\left[ L_{\nu}^{\alpha \beta} \Pi_{\alpha \beta}^{\rm Im} \right]^{\rm (INT)}  
&=& -4 q^2 \sum_{j} \left[ A_{I}^{i} \left( \Pi_{L}^{j} + \Pi_{T}^{j} 
\right) 
+ B_{\rm 1INT}^{j} \Pi_{T}^{j} 
\right. \nonumber \\ && \left. \mbox{}
+ B_{\rm 2 INT}^{j} \Pi_{A}^{J} 
+ C_{\rm INT}^{j} \Pi_{VA}^{j} 
\right],
\end{eqnarray}
where the sum over $j=p,n,e^-,\mu^-$ is understood.
The vector polarization tensor $\Pi_{\alpha \beta}^{{\rm Im} V}$ for each contribution can be 
represented by two independent components in the frame of 
$q^\mu \equiv  (q_0^{}, \abs{\bm{q}},0,0)$, which yields 
$\Pi_{T} = \Pi_{22}^{V} = \Pi_{33}^{V}$ and $\Pi_{L} = - (q^2 / \abs{\bm{q}}^2) 
\Pi_{00}^{V}$.
The axial vector and mixed pieces are found to be 
$\Pi_{\alpha \beta} ^{\rm Im (V-A)} (q) = i \epsilon_{0 \alpha \beta \eta}^{} q^{\eta} \Pi_{VA}$. 
The explicit forms of 
$\Pi_{T}$, $\Pi_{L}$, $\Pi_{VA}$, and $\Pi_{A}$  for the nucleon
are expressed as
\begin{eqnarray}
\Pi_{T} &=& \frac{1}{4\pi \abs{ \bm {q} }} \left[\ \left( M^{*2} 
+ \frac{q^4}{4 \abs{ \bm {q} }^2} + \frac{q^2}{2} \right) (E_F-E^{*}) 
\right. \nonumber \\ && \quad \left. \mbox{} 
+ \frac{q_0^{} q^2}{2 \abs{ \bm {q} }^2} ( E_F^2 - E^{*2} ) 
+ \frac{q^2}{3 \abs{ \bm {q} }^2} ( E_F^3 - E^{* 3} ) \right], 
\nonumber \\
\Pi_{L} &=& \frac{q^2}{2\pi \abs{ \bm {q} }^3} \left[ \  \frac{1}{4} ( E_F - E^{*} ) q^2 
\right. \nonumber \\ && \left. \quad \mbox{} 
+ \frac{q_0^{}}{2} ( E_F^2 - E^{*2} ) + \frac{1}{3 } ( E_F^3 - E^{* 3} )  \right],
\nonumber \\
\Pi_{VA} &=& \frac{iq^2}{8 \abs{ \bm {q} }^3} \left[ \left( E_F^2 - E^{* 2} \right) 
+ q_0^{} \left( E_F - E^{*}\right) \right],
\nonumber \\
\label{eq:model22}
\Pi_{A} &=& \frac{i M^{*2}}{2\pi \abs{ \bm {q} }} \left( E_F - E^{*}\right).
\end{eqnarray}
For leptons, we have similar expressions, but the lepton properties are assumed to be 
the same as those in free space.
Thus in Eq.~(\ref{eq:model22}) the in-medium quantities should be replaced by the corresponding
free space quantities in the case of lepton targets.
The functions appearing in Eq.~(\ref{eq:model18}) are given for the weak contributions,   
\begin{align}
\label{model22a}
A_{ W} &= \left( \frac{2E (E-q_0^{}) + \frac{1}{2} q^2}{\abs{ \bm {q} }^2} \right) 
\nonumber \\ & \quad \mbox{} \times
\left[ (F_1^{{ W}})^ 2 + G_A^{2 } - \frac{q^2}{4M^2} (F_2^{{ W}})^{2} \right],  
\nonumber \\
B_{1 W} &= \left[ (F_{1}^{ W})^{2} + G_A^{ 2} - \frac{(F_2^{ W})^{ 2} q^2}{4M^2} \right],  
\nonumber \\
B_{2 W} &= - \left[ G_A^{2} + \frac{q^2}{2m M} F_{1}^{W} F_2^{W} 
\right. \nonumber \\ & \left. \qquad \mbox{} 
- \frac{(F_2^{W})^{ 2} q^2}{4M^2} \left( 1 + {q^2}/{4m^2} \right) \right], 
\nonumber \\
C_{ W} &= -2  ( 2E - q_0^{}  ) \left[ F_1^{W} G_A^{W} + \frac{m}{M} F_2^{W} G_A \right],
\end{align}
while for the EM contributions they are given by 
\begin{align}
\label{model22b}
A_{\rm EM} &= \left[ \frac{2E \left( E - q_0^{} \right) 
+  q^2/2}{\abs{ \bm {q} }^2}\left( bq^2 -a \right) + \frac12 b q^2 
\right]   
\nonumber \\ & \quad \mbox{} \times
\left[ (F_{1}^{\rm EM})^{ 2} - \frac{(F_2^{\rm EM})^2  q^2}{4M^2}\right],  
\nonumber \\
B_{\rm 1EM} &= - \frac{1}{2} \left( bq^2 + a \right) \left[ ( F_{1}^{\rm EM} )^2 
- \frac{(F_{2}^{\rm EM})^2 q^2}{4 M^2} \right], 
\nonumber \\
B_{\rm 2EM}^{} &= \frac{1}{2} \left( bq^2 + a \right) \Biggr[ \frac{q^2}{2mM} F_{1}^{\rm EM} 
F_{2}^{\rm EM}  
\nonumber \\ &\mbox{}
- \frac{(F_2^{\rm EM})^{ 2} q^2}{4 M^2} \left(1+ \frac{q^2}{4m^2} \right) \Biggr], 
\end{align}
where
\begin{eqnarray}
a = 4 \left( f_{m \nu}^{2} + g_{1\nu}^2\right), \qquad
b = \frac{f_{2 \nu}^2 + g_{2\nu}^2}{m_{e}^2},
\end{eqnarray}
which are related to the charge radius and magnetic moment of the neutrino through 
Eqs.~(\ref{eq:model4})--(\ref{eq:model5}).
The interference terms are given obtained as
\begin{align}
\label{model22c}
A_{\rm INT}^{} &= c \left( \frac{2E \left( E - q_0^{} \right) 
+ \frac{1}{2} q^2 }{\abs{ \bm {q} }^2} \right) 
\nonumber \\ & \quad \mbox{}
\times \left[ F_{1}^{ W} F_{1}^{\rm EM} + \frac{q^2}{4M^2} F_{2}^{ W} F_{2}^{\rm EM} 
\right], \nonumber \\
B_{1\rm INT}^{} &= c \left[ F_{1}^{W} F_{1}^{\rm EM} + \frac{q^2}{4M^2} F_{2}^{ W} 
F_{2}^{EM} \right], \nonumber \\
B_{2\rm INT}^{} &= -c q^2 \Biggr[ \frac{F_{2}^{ W} F_{2}^{\rm EM}}{4 M^2} 
\left( 1 +  \frac{q^2}{4m^2} \right) 
\nonumber \\ & \mbox{}
- \frac{( F_{1}^{ W} F_{2}^{\rm EM} + F_{2}^{ W} F_{1}^{\rm EM} )}{4 mM} 
\Biggr], \nonumber \\
C_{\rm INT}^{} &= c \left( 2E -q_0^{}\right) \Biggr[ \frac{m}{M} F_{2}^{\rm EM} G_A^{} 
-  F_{1}^{\rm EM} G_A^{} \Biggr],
\end{align}
with $c = f_{m \nu} + g_{1 \nu}$.

In the present work, we consider only the NMFP of the neutrino elastic scattering, 
and do not consider that of neutrino absorption, because the NMFP of neutrino scattering
is dominated by the neutrino elastic scattering~\cite{SLVZ03}. 
The inverse mean free path of the neutrino is straightforwardly obtained 
by integrating the differential cross section of Eq.~(\ref{eq:model15}) over 
the energy transfer $q_0^{}$ and the three-momentum transfer $\abs{\bm{q}}$. 
The final expression for the NMFP as a function of the initial energy at a fixed baryon 
density can be obtained as~\cite{RPL97}
\begin{align}
\label{eq:model23}
\frac{1}{\lambda (E_\nu)} &= \int_{q_0^{}}^{2E_\nu - q_0^{}} d \abs{ \bm {q} } 
\int_0^{2E_\nu} d q_0^{} \frac{\abs{\bm{q}}}{E'_\nu E_\nu} \frac{2\pi}{V} 
\frac{d^3 \sigma}{d^2 \Omega' dE'_\nu},
\end{align}
where the final and initial neutrino energies are related as $E'_\nu = E + q_0^{}$.
More detailed explanations for the determination of the lower and upper limits of the integral 
can be found in Ref.~\cite{RPL97}.

\section{\boldmath Models for Matter} \label{mattermodels}

Since we are interested in the scatterings of neutrinos with the constituents of matter 
at zero temperature, the in-medium properties of nucleons are required as discussed in the previous section.
The interactions of the nucleon in matter are described by an effective chiral 
Lagrangian given as
\begin{equation}
\mathcal{L} = \mathcal{L}_{N} + \mathcal{L}_{M},
\label{eq:rmf1}
\end{equation}
where $\mathcal{L}_{N}$ is the interactions of the nucleon 
with mesons, and the mesonic Lagrangian $\mathcal{L}_{M}$
contains meson self-interactions.
For this purpose, we introduce the pion field through the chiral field $\xi$ as
\begin{equation}
\xi^2 = U = \exp(2i\pi/f_\pi),
\end{equation}
where $\pi = \frac12 \bm{\tau} \cdot \bm{\pi}$ and the pion decay constant is $f_\pi = 93$~MeV.
To develop chiral invariant interactions, we introduce the vector and axial 
vector currents, $v_\mu^{}$ and $a_\mu^{}$, as
\begin{eqnarray}
v_\mu^{} &=& - {\textstyle \frac{i}{2}} \left( {\xi}^{\dagger}\partial_\mu {\xi} 
+ {\xi} \partial_\mu {\xi}^{\dagger} \right) , 
\nonumber \\
a_\mu^{} &=& -{\textstyle \frac{i}{2}} \left( {\xi}^{\dagger}\partial_\mu {\xi} 
- {\xi} \partial_\mu {\xi}^{\dagger}\right) ,  
\end{eqnarray}
so that $v_\mu^{} = v_\mu^\dagger$ and $a_\mu^{} = a_\mu^\dagger$.
Following Refs.~\cite{FST96a,GM91,CRD96}, we write the effective Lagrangian  as~\cite{FST96a}
\begin{eqnarray}
\label{eq:rmf2}
\mathcal{L}_N &=& \bar{\psi} \Bigl[ i \gamma^\mu \left( \partial_\mu^{} 
+ i v_\mu^{} + i g_\rho^{} \rho_\mu^{} + i g_\omega^{} \omega_\mu^{} \right) 
\nonumber \\ && \mbox{} \quad
+ g_A^{} \gamma^\mu \gamma_5 a_\mu^{} - M_N^{} + g_\sigma^{} \sigma \Bigr] \psi ,
\end{eqnarray}
where $\psi$ is the nucleon isodoublet defined as
\begin{equation}
\psi = \begin{pmatrix} p \\ n \end{pmatrix}
\end{equation}
with $M_N^{}$ being the nucleon mass.
The Lagrangian $\mathcal{L}_N$ contains the interactions of the nucleon with the $\rho$ vector meson
($\rho_\mu^{} = \frac12 \bm{\tau} \cdot \bm{\rho}_\mu^{}$), $\omega$ vector meson ($\omega_\mu^{}$), 
and the scalar $\sigma$ meson ($\sigma$).

The Lagrangian $\mathcal{L}_{M}$ of the mesonic part reads~\cite{FST96a,GM91,CRD96}
\begin{eqnarray}
\label{eq:rmf4}
\mathcal{L}_{M} 
&=& \frac{f_\pi^2}{4} \,\mbox{Tr} \left( \partial_\mu U \partial^\mu U^\dagger \right) 
+ \frac{f_\pi^2 m_\pi^2}{4} \, \mbox{Tr} \left( U + U^\dagger -2 \right) 
\nonumber \\ && \mbox{}
+ \frac{1}{2} \partial_\mu^{} \sigma \partial^\mu \sigma 
- \frac{1}{2} \,\mbox{Tr} \left(\rho_{\mu \nu}^{} \rho^{\mu \nu} \right) 
- \frac{1}{4} \omega_{\mu \nu} \omega^{\mu \nu} 
\nonumber \\ && \mbox{}
+ \frac{1}{2} m_\omega^2 \, \omega_\mu \omega^\mu 
+ m_\rho^2 \, \mbox{Tr} \left( \rho_{\mu}^{} \rho^{\mu} \right) 
\nonumber \\ && \mbox{}
- \frac{b}{3} M_N \left( g_\sigma \sigma \right)^3 - \frac{c}{4} \left( g_\sigma \sigma \right)^4,
\end{eqnarray}
where 
\begin{eqnarray}
\omega_{\mu\nu}^{} = \partial_\mu^{} \omega_\nu^{} - \partial_\nu^{} \omega_\mu^{}, 
\qquad
\rho_{\mu\nu}^{} = \partial_\mu^{} \rho_\nu^{} - \partial_\nu^{} \rho_\mu^{}.
\end{eqnarray}
In the Hartree mean field approximation, the $\pi$ meson makes no contribution because of 
its negative intrinsic parity. 
Throughout the present calculation, we use $M_N^{} = 939$~MeV, $m_\rho^{} = 770$~MeV, 
$m_\omega^{} = 783$~MeV,
and $m_\sigma^{} = 520$~MeV.
We use the coupling constants determined in Refs.~\cite{GM91,CRD96}, i.e.,
$\left( g_\sigma^{} / m_\sigma^{} \right)^2 = 9.148~\mbox{fm}^2$,
$\left( g_\omega^{} / m_\omega^{} \right)^2 = 4.820~\mbox{fm}^2$,
$\left( g_\rho^{} / m_\rho^{} \right)^2 = 4.791~\mbox{fm}^2$, $b = 3.478 \times 10^{-3}$, 
and $c = 1.328 \times 10^{-2}$.
(We note that similar approaches were used successfully in Refs.~\cite{SHM05b,Reinhard89,SW86,Ring96,Serot92}.)

The Lagrangian for leptons reads
\begin{equation}
\label{eq:rmf5}
\mathcal{L}_{l} = \sum_{l = e^{-}, \, \mu^{-}, \nu_e, \nu_\mu} 
\bar{\psi}_l \left(\gamma^\mu \partial_\mu - m_l^{} \right) \psi_l
\end{equation}
where $m_l^{}$ denotes the lepton mass. 
In the present work, leptons are assumed to be free by forming the Fermi gas. 
Once the Lagrangian is given, we can obtain and solve the Euler-Lagrange equations
to compute the matter properties with the constraints of  the beta equilibrium,
which states the relation of the chemical potentials as
$\mu^{\rm chem.}_n + \mu^{\rm chem.}_{\nu_e} = \mu^{\rm chem.}_p + \mu^{\rm chem.}_e$,
and the charge neutrality,  $\rho_e^{} + \rho_\mu^{} = \rho_p^{}$.
The total baryon density is given by $\rho_B^{} = \rho_p^{} + \rho_n^{}$.

With the parameters given above, we obtain the saturation density $\rho_0^{} = 0.15~\mbox{fm}^{-3}$ 
with the binding energy $E/A  = 16.30~\mbox{MeV}$ at the saturation density.
The calculated compression modulus is $K= 219~\mbox{MeV}$, 
which is in good agreement with the empirical value of $K = 210 \pm 30$~MeV~\cite{Blaizot80}, 
and the symmetry energy coefficient $E_{\rm sym}$ is obtained to be around $32.5~\mbox{MeV}$. 
The effective nucleon mass $M_N^*$ is estimated to be $0.78\,M_N$.
These values are consistent with those obtained in the QMC model~\cite{Guichon88} which will be
used to calculate the in-medium nucleon form factors.

\section{Medium modification of the nucleon weak and electromagnetic form factors 
\label{medium}}

In order to estimate the effects of the in-medium modified nucleon 
form factors for neutrino scattering, 
we need to replace the free-space form factors, $G_A(0)$, $F_{2}^{ W}(0)$, and $F_{2}^{\rm EM}(0)$, 
in the formalism discussed in Sec.~\ref{scattering} by
the in-medium form factors, $G_A^*(0)$, $F_{2}^{ W*}(0)$, and $F_{2}^{\rm EM*}(0)$, respectively.
Since $F_1^{ W*} (0)$ and $F_{1}^{\rm EM*}(0)$ correspond to the  
weak-vector charge and the EM charge normalizations, respectively, their values do not 
change from the free space values.
Hereafter, quantities with an asterisk stand for those in nuclear medium.

For the estimates of the in-medium nucleon form factors, we make use of the QMC
model~\cite{STT05,LTTWS98b,LTT01,Thomas14,TKS03}. 
The QMC model~\cite{Guichon88} has been successfully applied to many problems  
of nuclear physics and hadron properties in nuclear medium. 
Some details of the applications can be found, for example, in 
Refs.~\cite{STT05,LTTWS98b,LTT01,Thomas14,TKS03,TSS04,TST00}. 
In the earlier works of neutrino-nucleus interaction of Refs.~\cite{TKS03,CCKSKTM13}, 
the in-medium modification of the nucleon weak form factors was applied, 
but the effect of the in-medium EM form factors is yet to be explored. 
Certainly, one can expect that the in-medium nucleon EM form factors can influence the beta 
equilibrated matter composition as well as the neutrino-matter scattering cross sections.
These are what we pursue in the present work.

In the QMC model, the medium effects arise from the self-consistent exchange  
of the scalar ($\sigma$) and vector ($\omega$ and $\rho$) meson fields 
directly coupled to the confined quarks rather than to the point-like nucleon.
In the following we consider symmetric nuclear matter 
in the Hartree mean filed approximation.
This may be justified because, the differences 
in the results between the Hartree and Hartree-Fock 
calculations are found to be relatively small. 
In particular, the energy densities per nucleon for symmetric nuclear matter 
are nearly identical at the cost of complications introduced in the Hartree-Fock treatment. 
Thus the use of the Hartree approximation in this exploratory study would be enough. 
More details on the Fock terms in the QMC model can be found in Ref.~\cite{KTT98}.
We consider the rest frame of the symmetric nuclear matter in the following to be
consistent with Sec.~\ref{mattermodels}.
The effective Lagrangian for a symmetric nuclear matter is given by~\cite{STT05,STT96}
\begin{eqnarray}
  \label{eqintro1}
  \mathcal{L}_{\rm QMC} & = & \bar{\psi} \left[ i\gamma \cdot \partial - M_{N}^{*}({\sigma}) - g_\omega^{}
  {\omega}^{\mu} \gamma_\mu \right] \psi 
\nonumber \\ && \mbox{}
+ \mathcal{L}_\textrm{meson},
\end{eqnarray}
where $\psi$, ${\sigma}$ and ${\omega}$ are respectively the nucleon, scalar $\sigma$, 
and vector $\omega$ fields, and the effective nucleon mass 
$M_{N}^{*} \left(\hat{\sigma} \right)$ is defined by 
\begin{align}
  \label{eqintro2}
M_{N}^{*} \left({\sigma} \right) &\equiv M_{N}^{} - g_\sigma(\sigma)  {\sigma},
\end{align}
with $g_{\sigma}(\sigma)$ and $g_{\omega}$ being respectively the $\sigma$-dependent 
nucleon-$\sigma$, and nucleon-$\omega$ coupling constants. 
Because in symmetric nuclear matter the isospin-dependent $\rho$-meson filed vanishes in the 
Hartree approximation, we do not explicitly include the $\rho$ meson.
The free meson Lagrangian density in Eq.~(\ref{eqintro1}) is defined by
\begin{eqnarray}
  \label{eqintro3}
  \mathcal{L}_\textrm{meson} &=& \frac{1}{2} (\partial_\mu {\sigma} \partial^\mu {\sigma} -
  m_\sigma^2 {\sigma}^2) - \frac{1}{2} \partial_\mu {\omega}_\nu 
  (\partial^\mu {\omega}^\nu - \partial^\nu {\omega}^\mu)
  \nonumber \\
  &&\mbox{} + \frac{1}{2} m_\omega^2 \omega^\mu \omega_\mu . 
\end{eqnarray}

In the mean-field approach, the nucleon Fermi momentum $k_F$ and the scalar density $\rho_{s}^{}$ 
in symmetric nuclear matter are defined as
\begin{align}
  \label{eqintro4}
  \rho_{B}^{} &= \frac{\gamma}{(2\pi)^3} \int d \bm{k}\, \theta ( k_F^{} - | \bm{k} | ) = 
  \frac{\gamma \, k_F^3}{3 \pi^2}, \nonumber \\
  \rho_{s}^{} &= \frac{\gamma}{(2\pi)^3} \int d \bm{k} \, \theta ( k_F^{} - | \bm{k} | )
  \frac{M_N^{*} (\sigma)}{\sqrt{M_N^{*2} (\sigma ) + \bm{k}^2}},
\end{align}
where $\gamma =$ 4 for symmetric nuclear matter.
For asymmetric nuclear matter $\gamma = 2$, and the Fermi momenta of the proton and neutron $k_{F}^{p,n}$ 
are determined by $\rho_p^{}$ and $\rho_n^{}$, respectively, with $\rho_B^{} = \rho_p^{} + \rho_n^{}$.

In the QMC model~\cite{STT05,STT96}, the nuclear matter is treated as a collection of nucleons that are
assumed to be non-overlapping MIT bags~\cite{CJJTW74}. 
The Dirac equations for light quarks ($q = u$ or $d$) in the bag are given by
\begin{eqnarray}
&& \left[ i \gamma \cdot \partial_{x} - \left( m_q - V_{\sigma}^{q} \right) \mp \gamma^{0} \left( 
  V_{\omega}^{q} + \frac{1}{2} V_{\rho}^{q} \right) \right] \left( \begin{array}{c} \psi_u(x)  \\ 
  \psi_{\bar{u}}(x) \\ \end{array} \right) = 0 \nonumber \\
&& \left[ i \gamma \cdot \partial_{x} - \left( m_q - V_{\sigma}^{q} \right) \mp \gamma^{0} \left( 
  V_{\omega}^{q} - \frac{1}{2} V_{\rho}^{q} \right) \right] \left( \begin{array}{c} \psi_d(x)  \\ 
  \psi_{\bar{d}}(x) \\ \end{array} \right) = 0
\nonumber \\
    \label{eqintro5}
\end{eqnarray}
where the effective quark mass $m_q^{*}$ is defined as
\begin{align}
  \label{eqintro5a}
  m_q^{*} & \equiv m_q - V_{\sigma}^{q},
\end{align}
with $m_q$ being the light-quark current mass ($q=u,d$) 
and $V_\sigma^{q}$ the scalar potential.
In this model, we assume the SU(2) isospin symmetry for the light quarks, which gives $m_q = m_u = m_d$. 
In symmetric nuclear 
matter using the Hartree approximation, the isospin dependent $\rho$-meson mean field yields 
$V_\rho^{q} =$ 0 in Eq.~(\ref{eqintro5}). 
The scalar and vector mean field potentials in nuclear matter are defined as
\begin{equation}
  \label{eq:potqqmc}
V_{\sigma}^{q}  \equiv g_{\sigma}^{q} \sigma = g_{\sigma}^{q} \braket{\sigma}, \qquad
V_{\omega}^{q} \equiv g_{\omega}^{q} \omega = g_{\omega}^{q} \delta^{\mu 0} \braket{ \omega^{\mu} }.
\end{equation}

The bag radius of the hadron $h$ in nuclear medium $R_h^{*}$ is determined 
from the hadron mass stability 
condition against the variation of the bag radius (to be shown below), 
and the eigen-energies in units of $1/R_h^{*}$ are
\begin{align}
  \label{eq:pionmed9}
  \left( \begin{array}{c}
    \epsilon_u \\
    \epsilon_{\bar{u}}
  \end{array} \right)
  & = \Omega_q^* \pm R_h^* \left(  V^q_\omega + \frac{1}{2} V^q_\rho \right), \nonumber \\
  \left( \begin{array}{c} \epsilon_d \\
    \epsilon_{\bar{d}}
  \end{array} \right)
  &= \Omega_q^* \pm R_h^* \left(
  V^q_\omega
  - \frac{1}{2} V^q_\rho \right).
\end{align}
The effective mass of a hadron in nuclear medium $m_h^{*}$ is calculated as
\begin{align}
  \label{eq:pionmed10}
  m_h^{*} &= \sum_{j = q, \bar{q}} \frac{n_j \Omega_j^{*} -z_h^{}}{R^{*}_h} 
  + \frac{4}{3} \pi  R_h^{* 3} B,
\end{align}
while the in-medium bag radius $R^*_h$is determined by the conidtion,
\begin{align}
\left. \frac{\partial m_h^{*}}{\partial R_h} \right\vert_{R_h = R_h^{*}} = 0,
\end{align}
where $\Omega^{*}_q = \Omega^{*}_{\bar{q}} = \left[ x_q^2 + \left(R_h^{*} m_q^{*} \right)^2 \right]^{1/2}$, 
$m_q^{*} = m_q - g_\sigma^{q} \sigma = m_q - V^q_\sigma$.
In Eq.~(\ref{eq:pionmed10}), $z_h^{}$ is related with the bag-model quantity, 
which is determined by the hadron mass in free space
and the bag pressure $B = {\rm (170\, MeV)}^4$ that is fixed by 
the standard QMC model input for the nucleon in vacuum, $R_N = 0.8$ fm
with $m_q = 5$ MeV~\cite{STT05,STT96}.
For the quarks inside the bag of the hadron $h$, the lowest positive eigenfunctions of the bag satisfy the boundary 
condition at the bag surface, $j_0 (x_{q}) =  \beta_{q}\, j_1 (x_{q})$, where 
\begin{equation}
\beta_{q} = 
\sqrt{\frac{\Omega^*_{q} -(m^*_{q} R^*_h)}{\Omega^*_{q} + (m^*_{q} R^*_h)}},
\end{equation} 
with $j_0$ and $j_1$ being the spherical Bessel functions.

The scalar $\sigma$ and vector $\omega$ meson mean fields at the nucleon level can be related as
\begin{eqnarray}
  \label{eq:pionmed11}
  \omega &=& \frac{g_\omega^{} \rho_B^{} }{m_\omega^2},\\
  \label{eq:sigma}
  \sigma &=& \frac{4 g_\sigma^{}  C_N (\sigma)}{(2\pi)^3m_\sigma^2} \int d \bm{k} \,
  \theta (k_F^{} - | \bm{k}| ) \frac{M_N^{*}(\sigma)}{\sqrt{M_N^{*2}(\sigma) + \bm{k}^2}}, 
\end{eqnarray}
where $C_N (\sigma)$ is defined as
\begin{eqnarray}
  C_N (\sigma) = \frac{-1}{g_\sigma (\sigma =0)} \left[ \frac{\partial M_N^{*} (\sigma )}{\partial 
  \sigma } \right], 
\end{eqnarray}
For the point-like nucleon the value of $C_N (\sigma)$ is unity. 
The $C_N (\sigma)$ and the $\sigma$-dependent coupling $g_\sigma (\sigma)$ are the origin of the novel 
saturation properties achieved in the QMC model, and contain the dynamics of quark structure of the 
nucleon.
The quark structure of the nucleon is reflected in the effective nucleon mass $M_N^{*} (\sigma$) through 
a self-consistent way.
By solving the self-consistent equation for the scalar $\sigma$ mean field 
in Eq.~(\ref{eq:sigma}), the total energy per nucleon is obtained as
\begin{eqnarray}
  \label{eq:pionmed12}
  E^{\rm tot}/A &=& \frac{4}{(2\pi)^3 \rho_B^{}} \int d\bm{k} \, \theta (k_F^{} - | \bm{k} |) 
  \sqrt{M_N^{*2} (\sigma) + \bm{k}^2} 
  \nonumber \\ && \mbox{}
  + \frac{m_\sigma^2 \sigma^2}{2\rho_B^{}} + \frac{g_\omega^2 \rho_B^{}}{2 m_\omega^2}.
\end{eqnarray}

\begin{figure}[t]
\centering\includegraphics[width=0.95\columnwidth]{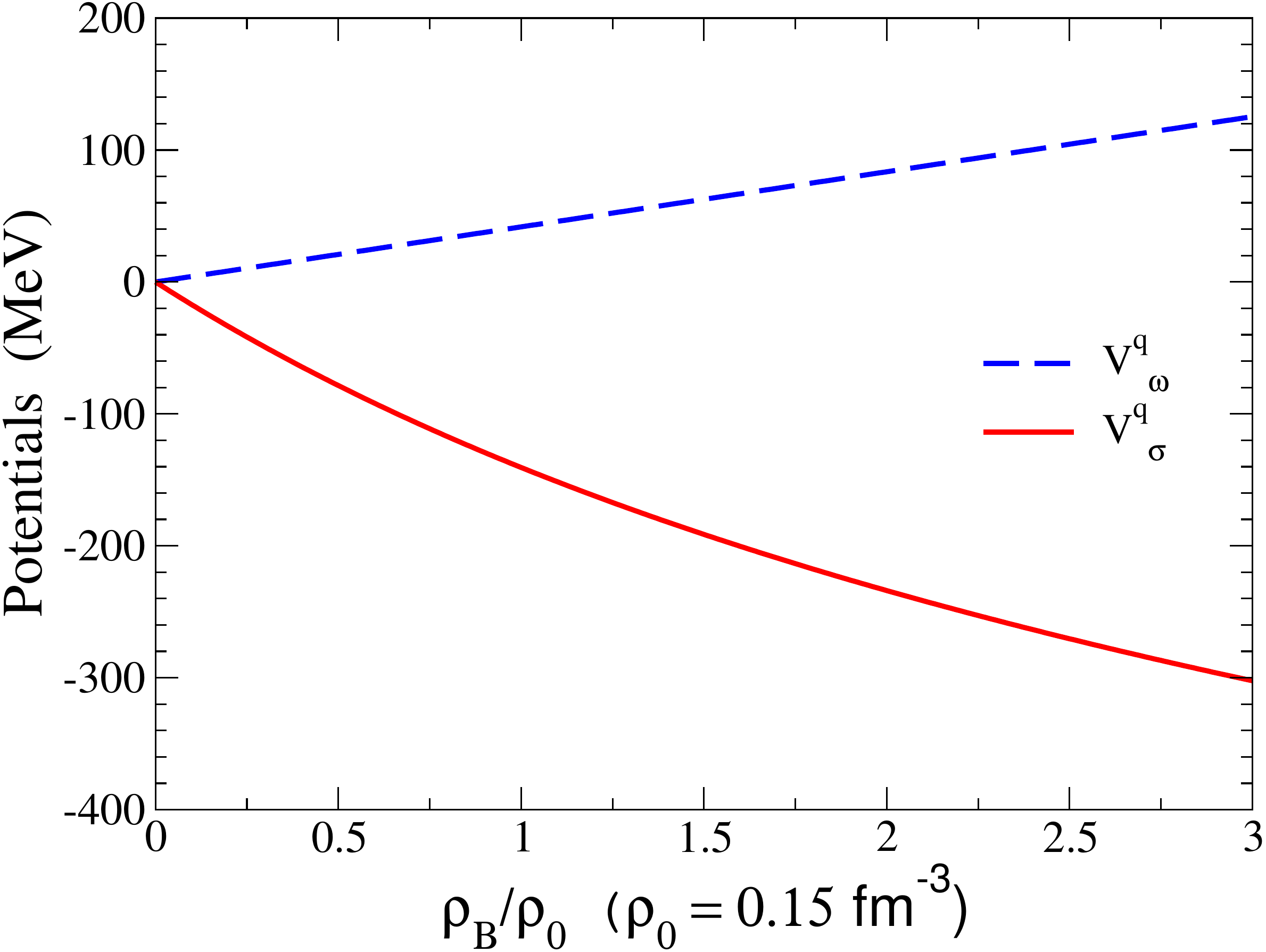}
\caption{\label{fig1} 
Quark potentials for $V^q_{\sigma}$ (solid line) and $V^q_{\omega}$ (dashed line) 
in symmetric nuclear matter for the current quark mass $m_q = 5~\mbox{MeV}$.}
\end{figure}

The coupling constants $g_\sigma$ and $g_\omega$ in Eq.~(\ref{eq:pionmed12}) are determined by 
fitting the binding energy of $15.7~\textrm{MeV}$ at the saturation density for symmetric nuclear 
matter. 
Results for the scalar and vector meson  coupling constants, incompresibility, and symmetry energy are 
listed in 
Table~\ref{tab:model2}. 
The resulting quark potentials of Eq.~(\ref{eq:potqqmc}) are shown in Fig.~\ref{fig1}.

\begin{table}[b]
\caption{Results for the coupling constants and the symmetric nuclear matter properties 
at saturation density $\rho_0^{} = 0.15~\mbox{fm}^{-3}$. 
The current quark mass ($m_q$), effective nucleon mass ($m_N^{*}$), the nuclear incompressibility 
($K$),
and the nuclear symmetry energy ($E_{\rm sym}$) are in units of MeV. 
We use $R_N = 0.8~\mbox{fm}$ (input) which is a standard value of the free nucleon bag radius 
in the QMC model.}
\label{tab:model2}
\addtolength{\tabcolsep}{5.8pt}
\begin{tabular}{cccccc} 
\hline \hline
$m_q$ & $g_{\sigma}^2 / 4 \pi$ & $g_{\omega}^2 / 4 \pi$ & $m_N^{*}$ & $K$ & $E_{\rm sym}$ \\[0.2em] 
\hline
$5.0 $ & $5.393$ & $5.305$ & $754.5$ & $279.3$ &  $35.0$
\\ \hline \hline
\end{tabular}
\end{table}

Referring the further details of the QMC model to Refs.~\cite{Guichon88,ST94,STT05}, 
we now discuss how the nucleon form factors in medium are computed based on the QMC model.
Using the $G$-parity argument (i.e., no second-class current), 
the charged weak-interaction vector and axial form factors 
for free-space nucleons are defined by
\begin{align}
\label{eq:qmcmed1}
& \braket{ p' s' \mid V_a^\mu (0) \mid p\, s } 
\nonumber\\
& = \bar{u}_{s'} (p') \left( F_{1}^{ W} (q^2) \gamma^\mu      
+ i \frac{F_2^{ W} (q^2)}{2M_N} \sigma^{\mu \nu} q_\nu^{} \right) \frac{\tau_a^{}}{2} u_{s} (p),  
\nonumber \\
& \braket{ p' s' \mid A_a^\mu (0) \mid p\, s }
\nonumber\\
& = \bar{u}_{s'} (p') \left( G_{A} (q^2) \gamma^\mu  \gamma_5^{}  
+  \frac{G_P (q^2)}{2M_N} q^\mu \gamma_5^{} \right) \frac{\tau_a^{}}{2} u_{s} (p),  
\end{align}
where $q_\mu^{} = p_\mu^\prime - p_\mu^{}$, and $\tau_a^{}$ $(a=1,2,3)$ are the Pauli matrices. 
The weak-vector form factors $F_1^{ W} (q^2)$ and $F_2^{ W} (q^2)$ are respectively assumed 
to be the same as those of the EM form factors $F_1^{\rm EM}(q^2)$ and $F_2^{\rm EM}(q^2)$, 
since they are the components of the corresponding isovectors.
Then, $F_1^{W, \rm EM} (q^2)$ and $F_2^{W, \rm EM} (q^2)$ are related to the electric $G_E (q^2)$ 
and magnetic $G_M (q^2)$ Sachs form factors as 
\begin{align}
G_E(q^2) &= F_1^{W, \rm EM}(q^2) + \frac{q^2}{4 M_N^2} F_2^{W, \rm EM}(q^2),
\nonumber \\ 
G_M(q^2) &= F_1^{W, \rm EM}(q^2) + F_2^{W, \rm EM}(q^2).
\end{align}

The induced pseudoscalar form factor $G_P(q^2)$ is dominated by the pion pole and can be
calculated using the PCAC relation~\cite{Thomas84,Miller84}. 
Nevertheless, the contribution from $G_P (q^2)$ to the cross section is proportional to 
$\mbox{(lepton mass)}^2/M_N^2$. 
Therefore, this contribution is small and will be neglected throughout the present calculation.

For estimating the in-medium modification of the nucleon form factors, 
we consider the rest frame of nuclear medium.
The in-medium nucleon form factors are calculated in the Breit frame~\cite{LTT01,LTTWS98} 
using the improved cloudy bag model (ICBM) of Ref.~\cite{LTW97} and the QMC model, which gives
\begin{align}
\label{eq:qmcmed2}
G_{E,M,A}^{\rm *\, QMC} (q^2) &=  \left( \frac{M_N^{*}}{E_N^{*}} \right)^2 G_{E,M,A}^{\rm *\, sph} 
( \tilde{q}^2 ), 
\end{align}
where $\tilde{q}^2 = (M_N^*/E_N^*)^2 q^2$ and the subscripts of $E$, $M$, $A$ mean electric, 
magnetic, and axial-vector form factors, respectively. 
The multiplication factor $(M_N^*/E_N^*)^2$ originates from the Lorentz contraction.
The in-medium nucleon energy is defined as $E_N^{*} = \sqrt{M_N^{*2}(\sigma) + \bm{q}^{2}/4}$,  
and $G_{E,M,A}^{\rm* \, sph} (q^2)$ are the form factors calculated with the 
static spherical MIT bag quark wave functions of nucleons~\cite{LTW97},   
applied in nuclear medium using the in-medium properties/inputs calculated by the QMC model. 
The ICBM includes a Peierls-Thouless projection to account for center-of-mass and recoil 
corrections as well as the Lorentz contraction of the internal quark wave function.
Note that the operators calculated in the QMC model are one-body nucleon operators
acting on the nucleons with the in-medium modified quark wave functions that reflect the changes of
the internal structure of the nucleon.
Therefore, many-body operators including two-body operators arising from meson exchange
are not explicitly included, although they can add extra contributions.

\begin{figure}[t]
\centering\includegraphics[width=0.95\columnwidth]{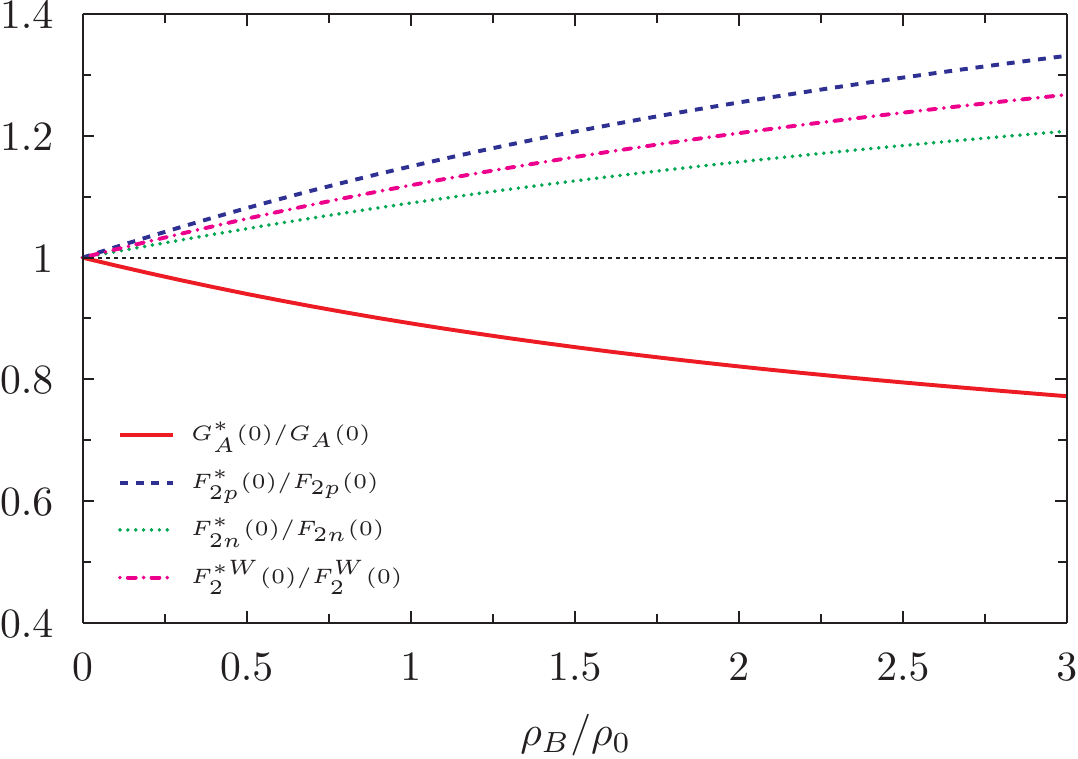}
\caption{\label{fig2} 
Ratios of the in-medium to free-space nucleon weak and EM form 
factors at $q^2 = 0$ versus $\rho_B^{}/ \rho_0^{}$ with $\rho_0 = 0.15~\mbox{fm}^{-3}$.
}
\end{figure}

The ratios of the in-medium to free-space nucleon form factors 
$G_{E,M,A}^{\rm *\, QMC} / (G_{E,M,A}^{\rm ICBM})_{\rm free}^{}$ 
are then calculated so that the in-medium nucleon form factors can be estimated. 
We use these ratios to get the in-medium nucleon form factors 
using the empirical form factors extracted in free space.
By making use of the empirical parameterizations in free space $G_{E,M,A}^{\rm emp}$, 
the in-medium nucleon form factors $G_{E,M,A}^{*} (q^2)$ can be estimated by  
\begin{align}
\label{eq:qmcmed3}
G_{E,M,A}^{*} (q^2) &= 
\left[ \frac{G_{E,M,A}^{\rm* \, QMC} (q^2)}{(G_{E,M,A}^{\rm ICMB})_{\rm free}^{} (q^2)} 
\right] G_{E,M,A}^{\rm emp} (q^2).
\end{align}
Note that the pion cloud effect is not taken into account in calculating the axial-vector form factor.
However, the normalized $q^2$ dependence reproduces relatively well the empirical 
parameterization in free space~\cite{LTT01}. 
Furthermore, the relative modification of $G_A^{*} (q^2)$ due to the pion cloud is 
expected to be small because the pion cloud contribution to $g_A^{}$, which is $G_A (q^2 =0)$, 
without specific center-of-mass corrections~\cite{Thomas84,Miller84}, is small.  
For our numerical calculations, as already mentioned, 
we use the current quark mass $m_u = m_d = 5~\mbox{MeV}$ 
assuming the SU(2) isospin symmetry and the free-space nucleon bag radius 
$R_N = 0.8~\mbox{fm}$. 
These values are considered to be standard in the QMC model~\cite{Guichon88,STT05,ST94,STT96}. 
The ratios of the in-medium to free-space nucleon form factors are then obtained
as shown in Fig.~\ref{fig2} as functions of $\rho_B^{}/ \rho_0^{}$.
It would be worthwhile to emphasize again that the form factor ratios 
presented in Fig.~\ref{fig2} are calculated based on the quark substructure of nucleons.
It should be noted that the counter effect of the weak magnetism $F_2^W$~\cite{HP03}, 
which makes the NMFP shorter by enhancing cross sections, turns out to give  
much smaller contribution than that of $G_A$.

We also confirmed that, contrary to the low $Q^2$ region of our interests, the ratio of the 
in-medium to free-space nucleon axial-vector form factors at high $Q^2$ does not show significant quenching.
In fact, the QMC model predicts small quenching even at $Q^2 = 2~\mbox{GeV}^2$~\cite{LTT01} and it becomes
almost negligible at $Q^2 = 3~\mbox{GeV}^2$.
However, since $G_A(Q^2)$ at large $Q^2$ is much smaller than that at small $Q^2$, the scattering cross sections
are dominated by the kinematic region of low $Q^2$.

\begin{figure*}[t]
\centering\includegraphics[width=0.97\textwidth]{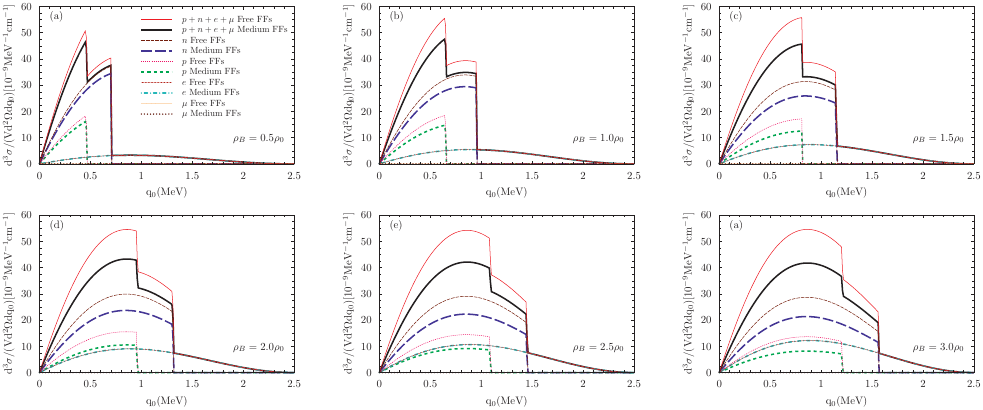}
\caption{ \label{fig3} (Colour online)
Differential cross sections of neutrino scatterings with the constituents of matter
as functions of $q_0^{}$ at the three-momentum transfer $\abs{\bm{q}} = 2.5~\mbox{MeV}$,  
and $E_\nu = 5~\mbox{MeV}$ for (a) $\rho_B^{} = 0.5\, \rho_0^{}$, (b) $\rho_B^{} = 1.0\, \rho_0^{}$,
(c) $\rho_B^{} = 1.5\, \rho_0^{}$, (d) $\rho_B^{} = 2.0\, \rho_0^{}$,
(e) $\rho_B^{} = 2.5\, \rho_0^{}$, and (f) $\rho_B^{} = 3.0\, \rho_0^{}$.
The thick-solid, thick-long-dashed, thick-short-dashed, thick-dashed-dotted and thick-dotted 
lines are respectively the differential cross sections for total $p+n+e+\mu$, 
neutron $n$, proton $p$, electron $e$, and muon $\mu$ obtained 
with the in-medium nucleon form factors, 
while the corresponding thiner-lines are those obtained with the free-space  
nucleon form factors. Note that, for the electron $e$ and muon $\mu$ cases, the lines 
are nearly degenerate for the results obtained with the in-medium nucleon form factors 
and those obtained with the free-space ones, and difficult to distinguish.
}
\end{figure*}

\section{Numerical results and discussions \label{numerical}}

We calculate the differential cross sections of neutrino scatterings 
with the constituents of matter at zero temperature 
as functions of the energy transfer $q_0^{}$ 
at $\abs{\bm{q}} = 2.5~\mbox{MeV}$ with the initial neutrino energy $E_\nu = 5~\mbox{MeV}$, 
which is the typical kinematics for the cooling phase of a neutron star~\cite{NBLM01}.
Our numerical results are shown in Fig.~\ref{fig3}, which
shows the total sum of the differential cross sections 
in vacuum (thin solid lines) and those in nuclear medium (thick solid lines)  
as well as the contributions from each target to the total sum of the differential 
cross sections (see the caption of Fig.~\ref{fig3} for details). 
In the present calculation, we set the charge radius of the neutrino $R_{V,A}=~0$ 
and neutrino magnetic moment 
$\mu_\nu = 0$ in order to focus on the different role of the nucleon form factors 
in vacuum and in medium. 
Also, when there is no neutrino trapping, the contribution of the neutrino form factors is 
too small compared to that of the nucleon form factors~\cite{HSM06}. 
Therefore, it does not give any visible effects in differential cross sections of 
neutrino scatterings, and we can safely neglect these effects 
in the estimation of the neutrino mean free path.

In Fig.~\ref{fig3}, we present the results for the neutrino scattering 
cross sections with $\rho_B^{} = 0.5\, \rho_0^{}$,
$\rho_B^{} = 1.0\, \rho_0^{}$, $\rho_B^{} = 1.5\, \rho_0^{}$, $\rho_B^{} = 2.0\, \rho_0^{}$, 
$\rho_B^{} = 2.5\, \rho_0^{}$,
and $\rho_B^{} = 3.0\, \rho_0^{}$.
In the case that the baryon density is lower than the normal nuclear matter 
density such as $\rho_B^{} = 0.5\, \rho_0^{}$,
the nucleon effective mass $M^{*}_N$ is not so much reduced from the free-space nucleon mass $M_N$,
namely, $M_N^{*} \approx M_N$. 
Consequently, this constrains the kinematic range of the energy transfer $q_0^{}$ for the nucleon. 
Since $q_0^{\rm max}$ is given by 
\begin{eqnarray}
q_0^{\rm max} &=& \sqrt{M_N^{*2} + (p_F^{} + \abs{\bm{q}})^2} - E_F 
\nonumber \\
&\simeq& 
\frac{1}{\sqrt{(M_N^*/p_F^{})^2 + 1}} \abs{\bm{q}},
\end{eqnarray}
the maximum value of $q_0^{}$ increases as
$M_N^*$ decreases at a given value of the Fermi momentum $p_F^{}$~\cite{RPL97,HW91}.
This kinematic cutoff is responsible for the sharp peak structure of the differential cross 
sections of the neutrino scattering with the target nucleon. 
At this low density, the fractions of the electron and muon scatterings with neutrinos are very small, 
and their masses are smaller than their Fermi momenta $p_F^{(\mu,e)}$, which
implies $q_0^{\rm max} \simeq \abs{\bm{q}}$. 
Therefore, the range of $q_0^{}$ is larger than that of nucleon targets.

The main effect of the in-medium nucleon weak and EM form factors is to suppress 
the neutrino-nucleon differential cross sections compared to those calculated 
with the free-space nucleon form factors. 
Because the weak interaction cross section has the dominant contributions 
from the vector ($F_1^W$) and (more dominant) axial-vector ($G_A$) form factors~\cite{GS84,IP82}, 
the quenching of $G_A$ in medium gives a larger impact on the reduction of 
the total differential cross section. 
As a result, the shape and magnitude of the differential cross section depend on 
the modification of the form factors. 
For baryon density $\rho_B^{} = 1.0\, \rho_0^{}$ and higher, the quenching of the axial-vector 
coupling constant gives more impacts and the differential cross sections of both 
the neutrino-neutron and neutrino-proton scatterings decrease accordingly.
The effect of the in-medium nucleon weak and EM form factors on the 
differential cross section is more pronounced at higher baryon density.
In addition, the range of the energy transfer $q_0^{}$ widens at higher density 
because the effective nucleon mass becomes smaller. 
Our results for $\rho_B^{} = 1.0\, \rho_0^{}$ are presented in Fig.~\ref{fig3}(b).
The results for baryon densities above the normal nuclear matter density are shown 
in Figs.~\ref{fig3}(c)-(f). 
One can verify that the impact of the in-medium nucleon weak and EM form factors 
is pronounced at higher densities.

\begin{figure}[t]
\centering\includegraphics[width=0.95\columnwidth]{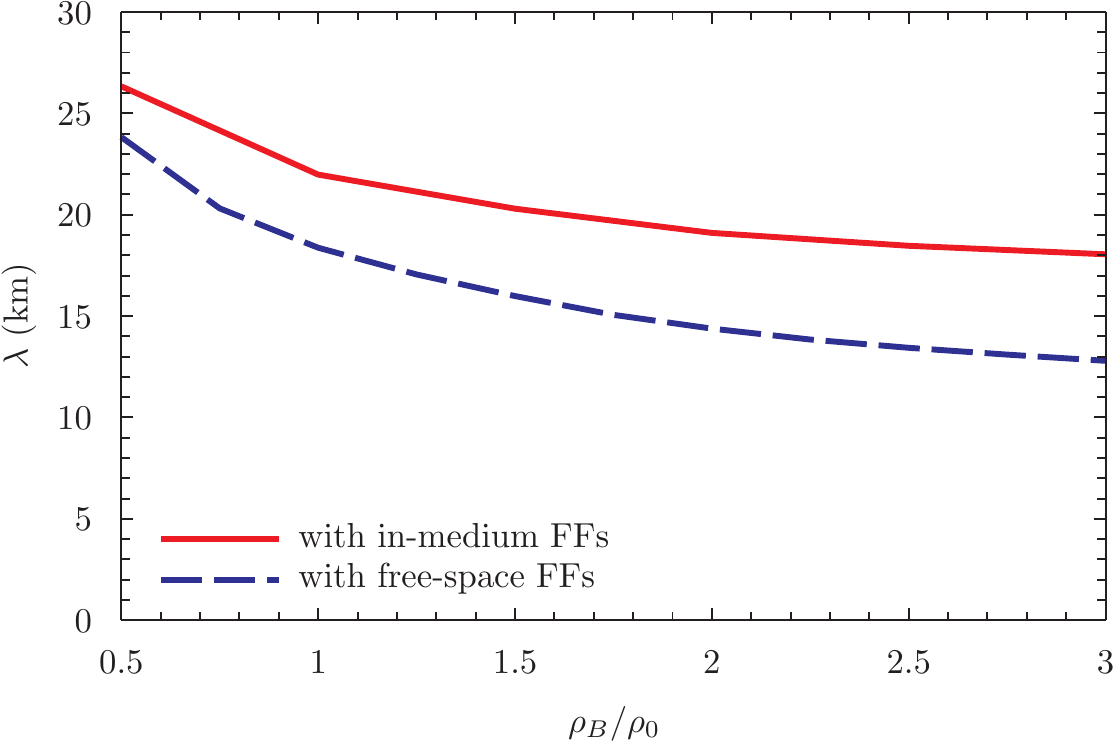}
\caption{\label{fig4} (Colour online)
The neutrino mean free path versus the nuclear density 
$\rho_B^{}/ \rho_0^{}$ at $\abs{\bm{q}} = 2.5~\mbox{MeV}$
and $E_\nu = 5~\mbox{MeV}$.}
\end{figure}

To study the impact of the in-medium nucleon weak and EM form factors, we calculate the NMFP  
for both cases with the vacuum and in-medium form factors. 
Figure~\ref{fig4} shows the NMFP with the free-space nucleon form factors (dashed line) 
and that with the in-medium modified nucleon form factors (solid line) 
at $\abs{\bm{q}} = 2.5~\mbox{MeV}$ and $E_\nu = 5~\mbox{MeV}$. 
As expected, the NMFP becomes longer by the in-medium modification of the
nucleon weak and EM form factors. 
Our results show that the in-medium modification of the nucleon form factors increases the NMFP by
10-40\%. 
This causes a faster cooling of neutron stars,  
since it makes easier for neutrinos to escape from the nuclear medium.

Our results show that the interactions between the neutrino with nucleons 
in nuclear medium becomes weaker than those in free space. 
Although the nucleon weak and EM form factors at $q^2 = 0$, i.e., $F_2^{ W}(0)$ and $F_2^{\rm EM}(0)$, 
respectively, are enhanced in nuclear medium, the quenched axial-vector coupling constant $G_A^*(0)$
gives a dominant contribution to reduce the cross section, which
results in the enhancement of NMFP.

\section{Summary \label{summary}}

To summarize, we have studied the impact of the in-medium modification of the nucleon weak and 
electromagnetic form factors on the neutrino scattering in the calculation of differential cross sections and 
the neutrino mean free path in dense matter using the results from a relativistic mean field model. 
The in-medium nucleon form factors are estimated by the quark-meson coupling 
model that is based on the quark degrees of freedom of the nucleon and nuclear matter
enjoying successful applications to describing the hadron and nuclear properties 
in nuclear medium.

The differential cross sections of the neutrino scatterings with the constituents of cold
matter were found to slowly decrease with increasing baryon density,
which results in the increase of the neutrino mean free path. 
This feature is sensitive to the in-medium modifications of 
the nucleon weak and electromagnetic 
form factors (in particular, that of the axial-vector form factor),  
and that the effect is pronounced for higher baryon densities.
The increases of the neutrino mean free path is estimated to be about 10-40\% 
compared with the results obtained with the free-space nucleon form factors.
This suggests that the cooling of a neutron star due to the escape 
of neutrinos is faster than the expectation based on the previous 
calculations using the free-space nucleon weak and electromagnetic form factors.
The impact of this observation on supernova explosion may be studied 
by developing more realistic simulations~\cite{MJBHMM15,MMHJ17} 
by including the medium effects.

As baryon density increases, hyperons such as $\Lambda$, $\Sigma$, and $\Xi$ are expected 
to appear in the nuclear medium~\cite{GM91,SM96,WCTTS13,MCS15,LLO17}. 
Since the interactions of the neutrino with hyperons are different 
from those with nucleons, it would be interesting
to investigate the effects of the in-medium weak and electromagnetic 
form factors of hyperons for estimating the neutrino mean free path. 
Inclusion of hyperon composition in matter and their in-medium form factors 
should give a more realistic description of the neutrino scattering 
with matter produced in neutron star and is expected to change 
the fractions of matter composition.

\begin{acknowledgments}
We are grateful to Hungchong Kim for fruitful discussions.
K.T. thanks the Asia Pacific Center for Theoretical Physics (APCTP) 
for warm hospitality and excellent supports during his visit and stay. 
P.T.P.H. was supported by the Young Scientist Training program of APCTP. 
The work of Y.O. was supported by the National Research Foundation of Korea under Grant 
No.~NRF-2015R1D1A1A01059603. 
The work of K.T. was supported by the Conselho Nacional de Desenvolvimento Cient\'{i}fico 
e Tecnol\'{o}gico - CNPq  
Grants, No.~400826/2014-3 and No.~308088/2015-8, 
and Funda{\c c}\~{a}o de Amparo \`{a} Pesquisa do Estado 
de S\~{a}o Paulo-FAPESP Grant, No.~2015/17234-0. The work of K.T.  
was also a part of the projects, Instituto Nacional de Ci\^{e}ncia e 
Tecnologia - Nuclear Physics and Applications 
(INCT-FNA), Brazil, Proc. No. 464898/2014-5, 
and FAPESP Tem\'{a}tico, Brazil, Proc. No. 2017/05660-0.
\end{acknowledgments}

\end{document}